\documentclass[fleqn, usenatbib]{mnras}

\usepackage{amssymb}
\usepackage{dcolumn}
\usepackage{bm}
\usepackage{graphicx}
\usepackage{aas_macros}
\usepackage{subfig}
\usepackage{amsmath}

\begin{document}
\label{firstpage}
\pagerange{\pageref{firstpage}--\pageref{lastpage}}

\title[GAMA Spheroid and Disk Mass Budget]{Galaxy And Mass Assembly (GAMA): the Stellar Mass Budget of Galaxy Spheroids and Disks}

\author[Moffett et al.]{Amanda J. Moffett,$^1$$^\dagger$ Rebecca Lange,$^1$ Simon P.~Driver,$^{1,2}$ Aaron S. G.
  \newauthor Robotham,$^1$ Lee S. Kelvin,$^3$ Mehmet Alpaslan,$^4$ Stephen K. Andrews,$^1$
  \newauthor Joss Bland-Hawthorn,$^5$ Sarah Brough,$^6$ Michelle E. Cluver,$^7$ Matthew
  \newauthor Colless,$^8$ Luke J. M. Davies,$^1$ Benne W. Holwerda,$^9$ Andrew M. Hopkins,$^6$
  \newauthor Prajwal R. Kafle,$^1$ Jochen Liske,$^{10}$ and Martin Meyer$^1$ \\
$^1$ICRAR, The University of Western Australia, 35 Stirling Highway, Crawley WA 6009, Australia\\
$^2$SUPA, School of Physics \& Astronomy, University of St Andrews, North Haugh, St Andrews, KY16 9SS, UK\\
$^3$Astrophysics Research Institute, Liverpool John Moores University, IC2, Liverpool Science Park, 146 Brownlow Hill, Liverpool, L3 5RF \\
$^4$NASA Ames Research Center, N232, Moffett Field, Mountain View, CA 94035, USA \\
$^5$Sydney Institute for Astronomy, School of Physics A28, University of Sydney, NSW 2006, Australia \\
$^6$Australian Astronomical Observatory, PO Box 915, North Ryde, NSW 1670, Australia \\
$^7$Department of Physics and Astronomy, University of the Western Cape, Robert Sobukwe Road, Bellville, 7535, South Africa \\
$^8$Research School of Astronomy and Astrophysics, Australian National University, Canberra, ACT 2611, Australia \\
$^9$University of Leiden, Sterrenwacht Leiden, Niels Bohrweg 2, NL-2333 CA Leiden, The Netherlands \\
$^{10}$Hamburger Sternwarte, Universitat Hamburg, Gojenbergsweg 112, 21029 Hamburg, Germany \\
$^\dagger${E-mail: amanda.moffett@uwa.edu.au} }

\date{Accepted XXX. Received YYY; in original form ZZZ}

\pubyear{2016}

\maketitle

\begin{abstract}
  We build on a recent photometric decomposition analysis of 7506 Galaxy and Mass Assembly (GAMA) survey galaxies to derive stellar mass function fits to individual spheroid and disk component populations down to a lower mass limit of log(M$_{*}$/M$_{\odot}$) $= 8$. We find that the spheroid/disk mass distributions for individual galaxy morphological types are well described by single Schechter function forms. We derive estimates of the total stellar mass densities in spheroids ($\rho_{spheroid} = 1.24\pm 0.49 \times 10^{8}$ M$_{\odot}$Mpc$^{-3}$h$_{0.7}$) and disks ($\rho_{disk} = 1.20\pm 0.45 \times 10^{8}$ M$_{\odot}$Mpc$^{-3}$h$_{0.7}$), which translates to approximately 50\% of the local stellar mass density in spheroids and 48\% in disks. The remaining stellar mass is found in the dwarf ``little blue spheroid'' class, which is not obviously similar in structure to either classical spheroid or disk populations. We also examine the variation of component mass ratios across galaxy mass and group halo mass regimes, finding the transition from spheroid to disk mass dominance occurs near galaxy stellar mass $\sim10^{11}$ M$_{\odot}$ and group halo mass $\sim10^{12.5}$ M$_{\odot}/h$. We further quantify the variation in spheroid-to-total mass ratio with group halo mass for central and satellite populations as well as the radial variation of this ratio within groups.

\end{abstract}


\setlength{\extrarowheight}{0pt}

\section{Introduction}

Spheroidal and disk galaxy structures are generally considered to result from separate formation mechanisms. In the simplest picture of galaxy structure growth, stellar spheroids arise from dissipationless accumulation of previously formed stars in mergers (e.g., \citealp{Cole2000}), and disks arise from star formation in the dissipational collapse of high angular momentum gas (e.g., \citealp{FE1980}). Thus, placing constraints on the balance of mass formed in spheroids and disks should tell us about the balance between the modes of galaxy formation that are dissipationless and dissipational. However, there are many potential complications to this simple picture. For example, it is now argued that galaxy bulges may initially form at high redshift from gas inflows enabled by disk instabilities and be further grown by minor mergers over time (e.g., \citealp{Parry2009}; \citealp{Hopkins2010}; \citealp{Bournaud2011}). The two-phase model of \citet{Driver2013} also envisions a transition from spheroid formation at z $\gtrsim$ 1.7, enabled primarily by major mergers, to disk formation at z $\lesssim$ 1.7, enabled primarily by gas accretion. Considering the role of accretion in more detail, a recent simulation analysis of \citet{Sales2012} suggests that the main influence on the mode of structure formation is actually the alignment of material accreted into the halo, where poorly aligned accretion events result in spheroid structures and well-aligned accretion events result in disk structures.

In galaxy formation simulations, reproduction of realistic galaxy bulge and disk structures has been a longstanding problem. Early issues with the overproduction of bulges in simulations have steadily improved (e.g., as recently reviewed by \citealp{SD2015}; \citealp{BC2016}). Hydrodynamical simulations are now able to produce even extreme bulgeless disk morphologies (e.g., \citealp{Governato2010}; \citealp{Brook2011}). However, it has been noted that similar bulge formation mechanisms lead to bulges that are still somewhat more massive than are typically observed for intermediate mass disk galaxies (see e.g., \citealp{Christensen2014}).

Hydrodynamical simulations of bulge formation have thus far been limited to relatively small samples of objects in zoom-in simulations, but cosmological semi-analytic models are now producing realistic galaxy spheroid \emph{populations} through merger-driven spheroid formation mechanisms, at least for intermediate to high mass galaxies (see \citealp{SD2015} and references therein). Moving forward with both types of models, accurate observational measurements of the galaxy mass assembled in spheroids and disks down to the low mass regime should provide important constraints on the ability of cosmological simulations to reproduce realistic structural properties for entire galaxy populations.

A number of authors have now produced measurements of the relative mass contribution of galaxy bulge and disk structures at low redshift, with the broad conclusion that galaxy stellar mass is nearly equally divided between spheroid and disk structures (e.g., \citealp{Driver2007full}; \citealp{Benson2007}; \citealp{Gadotti2009}). These studies differ subtly in the detailed mass breakdown, however, with estimated disk mass contributions ranging from 35-50\% (e.g., \citealp{Benson2007}; \citealp{Gadotti2009}; \citealp{Thanjavur2016}) or up to 59\% in the case of \citet{Driver2007full}.

In order to constrain spheroid and disk masses for large galaxy samples, measurements of this type rely on photometric decompositions of composite bulge and disk systems. A number of photometric structure decomposition codes have been developed for this purpose, including \textsf{GIM2D} \citep{GIM2D}, \textsf{BUDDA} \citep{BUDDA}, \textsf{GALFIT} \citep{GALFIT}, and \textsf{IMFIT} \citep{IMFIT}. Regardless of the decomposition routine used, understanding possible fitting systematics and estimating realistic parameter uncertainties is crucial to making an accurate estimate of the galaxy spheroid and disk mass budget in a large galaxy sample. In a recent analysis using \textsf{GALFIT}, \citet{Lange_decomp} address these issues by considering a large grid of structural fits with an array of initial guess parameters for each galaxy. The variety of initial model parameters guards against convergence to local rather than global minima in model fit parameters, and realistic systematic uncertainties are derived for fit parameters by quantifying the spread in the full ensemble of fit models for each galaxy.

In this work, we derive a new measurement of the stellar mass budget of galaxy spheroids and disks using the \citet{Lange_decomp} structural decomposition of Galaxy and Mass Assembly survey \citep{GAMAsurv,Driver2011} galaxies. We summarize our data and analysis methods in \S \ref{samp} and \S \ref{meth}. We then present separate bulge and disk stellar mass function fits for galaxies of various morphological types and derive estimates of the total galaxy stellar mass density of bulges and disks in \S \ref{res}, finding a nearly equal division been spheroid and disk mass in the local Universe. We also quantify the variation of spheroid-to-disk-mass ratio as a function of galaxy mass and group halo mass, finding that spheroid mass is only dominant at the highest galaxy and group halo mass scales. We briefly summarize and discuss these results further in \S \ref{conc}.

A standard cosmology of ($H_0$, $\Omega_m$, $\Omega_{\Lambda}$) = ($70$ km s$^{-1}$ Mpc$^{-1}$, 0.3, 0.7) is assumed throughout this paper, and h$_{0.7}$$=$$H_0$/($70$km s$^{-1}$ Mpc$^{-1}$) is used to indicate the $H_0$ dependence in key derived parameters.

\begin{figure}
\includegraphics[width=0.48\textwidth]{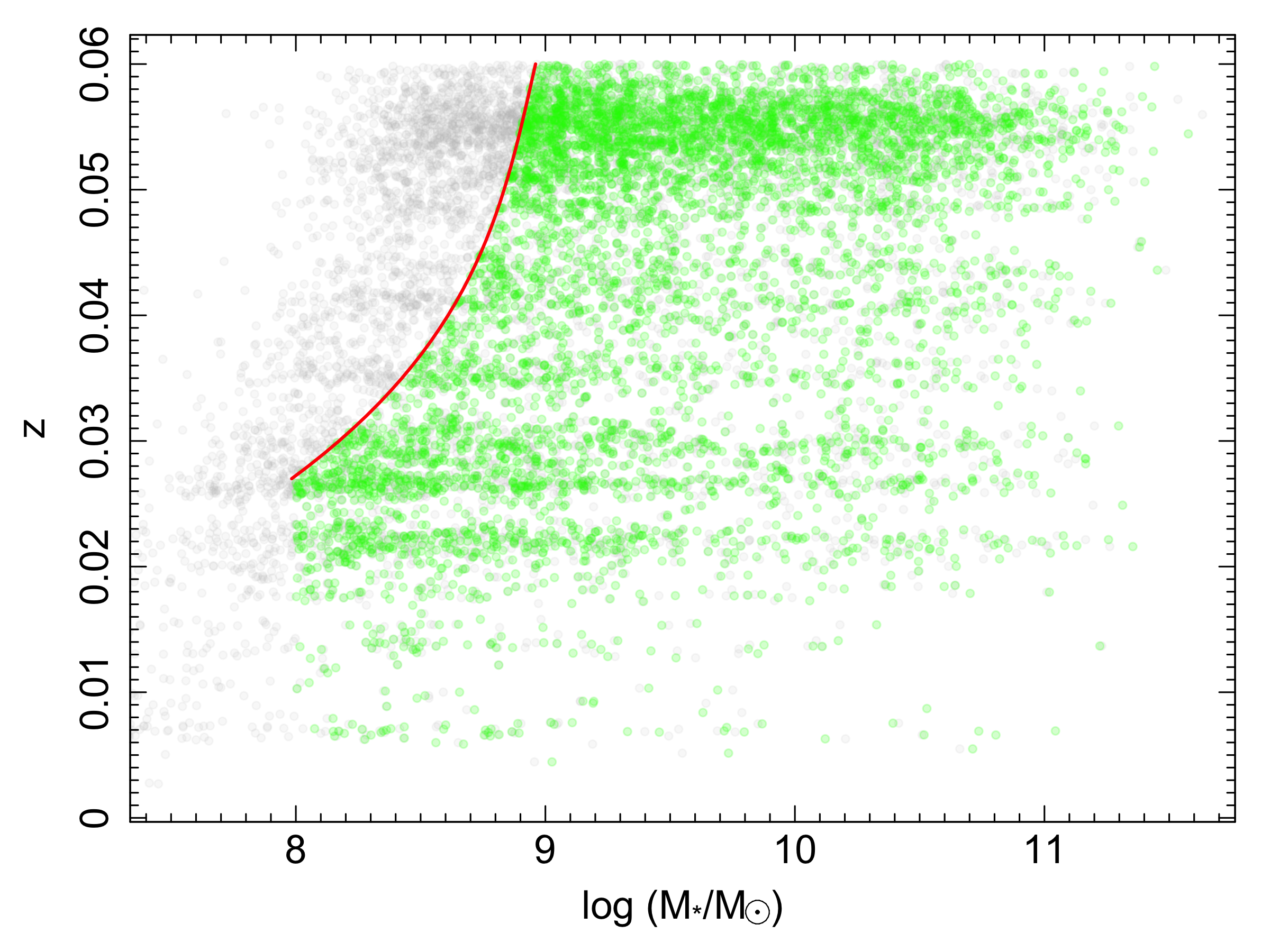}
\caption{The GAMA II structural decomposition sample in redshift versus stellar mass space (grey points), with green points indicating the sliding volume-limited subsample of galaxies we use to derive spheroid and disk stellar mass function fits. The red line indicates the mass limit as a function of redshift discussed in \S \ref{MLfits}.}
\label{fig:samp}
\end{figure}

\section{The GAMA II Structure Sample}
\label{samp}
Our data is taken from the Galaxy and Mass Assembly survey phase II, known as GAMA II.  GAMA is a combined spectroscopic and multi-wavelength imaging survey designed to study both galaxy-scale and large-scale structure (see \citealp{GAMAsurv,Driver2011} for an overview and \citealp{Hopkins_spec} for details of the spectroscopic data). The survey, after completion of phase II \citep{GAMADR2}, consists of three equatorial regions and two non-equatorial regions. The equatorial regions span approximately 5 deg in Dec and 12 deg in RA, centered in RA at approximately 9$^h$ (G09), 12$^h$ (G12) and 14.5$^h$ (G15). We use the three equatorial regions in this study, which are $>98$\% redshift complete to $r < 19.8$ mag \citep{GAMADR2} and combined total a sky area of 180 deg$^2$.

Within the GAMA equatorial regions, our structural fitting sample is derived from the GAMA II visual morphology catalog \citep{vismorph}, which contains $\sim$7500 objects from the GAMA tiling catalogue (TilingCatv44; \citealp{Baldry_tiling}) with survey\_class $\geq$ 1, extinction-corrected Sloan Digital Sky Survey (SDSS; \citealp{SDSScat}) $r$ band Petrosian magnitude of $r < 19.8$ mag, local flow-corrected redshift $0.002 < z < 0.06$, and normalized redshift quality nQ$>2$ (GAMA DistancesFramesv12; \citealp{Baldry2012}). These objects are visually classified into E, S0-Sa, SB0-SBa, Sab-Scd, SBab-SBcd, Sd-Irr, and “little blue spheroid” (LBS) galaxy types. We also judge 25 objects to be non-galaxy targets in the visual classification process \citep{vismorph}. These non-galaxy objects are subsequently omitted from our structural fitting sample.

\begin{figure*}
\includegraphics[width=0.9\textwidth]{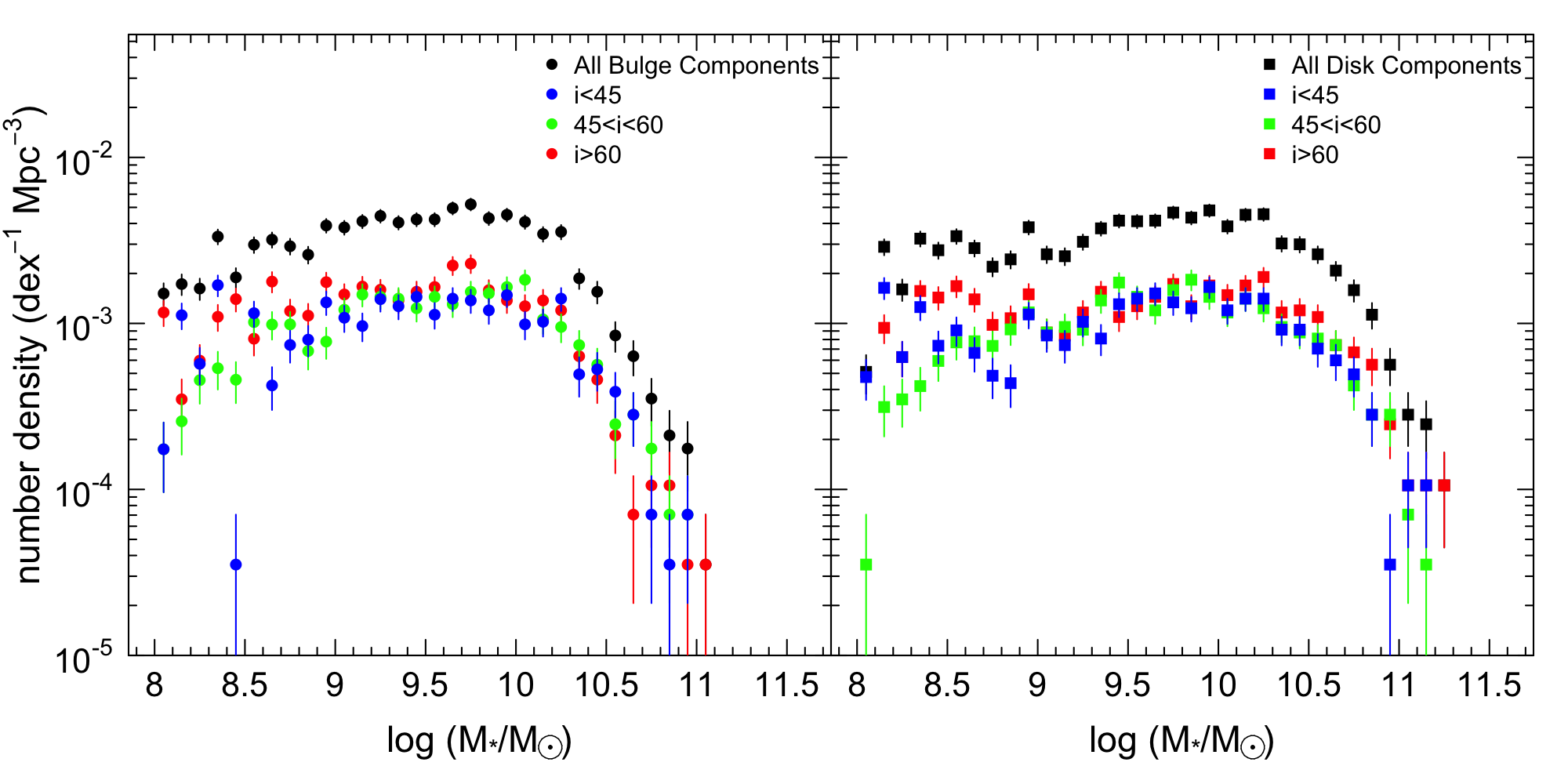}
\caption{Stellar mass functions for the bulge (left panel) and disk (right panel) components of our two-component (S0-Sa and Sab-Scd) systems. Each population has been divided into three separate, approximately equally sampled, inclination categories. No clear trend in stellar mass function shape as a function of inclination is observed.}
\label{fig:incltest}
\end{figure*}

\section{Methods}
\label{meth}
In this section, we briefly describe the procedure used for the GAMA II structural fitting analysis and our methods for deriving stellar mass function fits to the spheroid and disk populations.

\subsection{Structural Fitting and Decomposition}
\label{strucfits}
The structure sample of 7506 galaxies (see Fig. \ref{fig:samp}), excluding non-galaxy targets and one galaxy that is too large in angular size for effective analysis, has been fit in the SDSS $r$ band with two-dimensional \citet{Sersic1968} profile models. We use the structural fits described in detail by \citet{Lange_decomp}. This fitting procedure involves the use of \textsf{GALFIT} \citep{GALFIT} as implemented using the \textsf{SIGMA} wrapper code developed for GAMA by \citet{SIGMA}. \citet{Lange_decomp} takes a grid-based approach to structural fitting, defining a large grid of initial input parameters to guard against optimisation solutions that represent local rather than global minima. Further, rather than using a single best-fit model to infer structural parameters, we consider the full ensemble of ``good'' final model fits (as described fully by \citealp{Lange_decomp}) and define each structural parameter as the \emph{median} of the resulting model distribution. We also use these distributions to derive robust uncertainties on our model fit parameters.

Of the 7506 sample galaxies, 5259 have been morphologically classified as single-component systems, while 2247 have been classified as two-component systems. We use these morphological classifications to inform whether single-S\'ersic or double-S\'ersic models are most appropriate and default to model parameters derived from single-component fits for E, Sd-Irr, and LBS systems and double-component fits for S0-Sa and Sab-Scd systems. We note that barred galaxies identified in our visual morphology classification (types SB0-SBa and SBab-SBcd) are considered in the same category as their unbarred counterparts, as we find a low $\sim$12\% bar fraction in this sample. Since the most complicated models we fit in this analysis are effectively bulge plus disk models, it is likely that the central component masses for this small population of barred galaxies will actually reflect both bulge and bar masses.

\citet{Lange_decomp} identify a sample of two-component systems where the final derived fits include S\'ersic bulge n values that are smaller than the disc n values. These systems are flagged for exclusion in the derived component mass-size relation fits if they have disk n $>$ 2 or underestimated parameter uncertainties (100 S(B)0-S(B)a and 215 S(B)ab-S(B)cd galaxies excluded). These objects have a similar stellar mass distribution to their parent morphological type categories. We test whether or not the exclusion of these objects would alter the shape of our derived mass function fits and find that the S0-Sa bulge/disk mass function knee and slope parameters are consistent within estimated uncertainties whether these objects are included or excluded. The mass function shape parameters for separate Sab-Scd bulge and disk components differ slightly if these objects are excluded, however as we discuss further in \S \ref{func_comp} we find that single-S\'ersic fits are sufficient to describe this population and therefore do not include the Sab-Scd bulge plus disk fits in our final analysis. Since the exclusion of genuine objects would alter our mass function normalisation and the shape parameters of the mass functions we employ in our final analysis are not affected by the inclusion/exclusion of these objects, we elect to include them in our analysis.

Our structural fits for two-component systems yield bulge-to-total luminosity ratios, but we do not assume that these ratios translate directly to bulge-to-total mass ratios. Instead, we estimate the stellar mass contained in bulge and disk components separately, using the \citet{Taylor2011} calibration that relates optical colour ($g-i$) and mass-to-light ratio to stellar mass (see \citealp{Lange_decomp} for complete description). Briefly, we calculate this estimate by combining SDSS $r$-band bulge and disk magnitudes with $gri$ total and central PSF magnitudes measured using the \textsf{LAMBDAR} photometry code \citep{LAMBDAR}. We assume that PSF colours are equivalent to bulge colours and that bulge and disk fluxes sum to equal the total flux in each band. With these assumptions, we derive bulge and disk $g-i$ colours and $i$-band magnitudes, which we use to estimate component stellar masses according to the \citet{Taylor2011} relation:
\begin{equation}
\log{M_{*}/M_{\odot}} = -0.68 + 0.7~(g-i) - 0.4~(M_{i}-4.58) \text{.}
\end{equation}

For single-component galaxies, we use the total galaxy stellar mass estimates of \citet{Taylor2011} derived using GAMA optical photometry and stellar population synthesis modeling with a \citet{Chabrier} initial mass function. We include the additional mass scaling factors discussed by \citet{Taylor2011} that account for light missed in finite-size GAMA apertures by comparison to S\'ersic measures of total flux from \citet{SIGMA}.

As has been discussed frequently in the literature, it is important to consider how internal dust attenuation can alter not only the observed flux but also the structural parameters we infer from photometric data (e.g., \citealp{Byun1994}; \citealp{Evans1994}; \citealp{Mollenhoff2006}; \citealp{Gadotti2010}; \citealp{Bogdan2013a}). Particularly relevant to this analysis, \citet{Gadotti2010} and \citet{Bogdan2013b} found that dust effects can cause underestimation of both bulge n values and bulge-to-disk ratios. Further, \citet{Driver2007full} found that the $B$-band luminosity functions used to infer bulge and spheroid stellar mass densities required significant inclination-dependent corrections for such internal attenuation effects.

As a result of these concerns, we test whether or not our main products, the stellar mass functions of spheroids and disks, may require additional inclination-dependent corrections. First, considering the colours that are used to derive component mass estimates, we find that there is no overall trend between our measured component colours and the component axial ratios, implying that our colours are not affected by residual reddening in more edge-on objects. Further, we consider the mass functions we infer from both bulge and disk components of our two-component galaxies subdivided by inclination ranges. We estimate photometric inclination for each galaxy as $i=\cos^{-1}{\sqrt{ ((b/a)^{2} - q_{o}^{2})/(1-q_{o}^{2}) }}$ (where $b/a$ is the photometric axial ratio and the flattening parameter $q_{o}$ is assumed to be 0.2). We then split our sample into three broad inclination categories chosen to have approximately equal numbers in each category (see Fig. \ref{fig:incltest}). Examining the bulge and disk component mass functions, we find no obvious shift in the mass functions. Some small-scale differences in the three inclination categories can be seen, however the differences in the binned mass functions are in general comparable to the Poisson error bars on these points. As a result, we conclude that despite the fact that internal attenuation should affect structural measurements for individual galaxies in an inclination-dependent fashion, our mass functions averaged over entire populations appear to be insensitive to this effect, at least within the uncertainties implied by our sample and survey size.

\begin{figure*}
\includegraphics[width=1.0\textwidth]{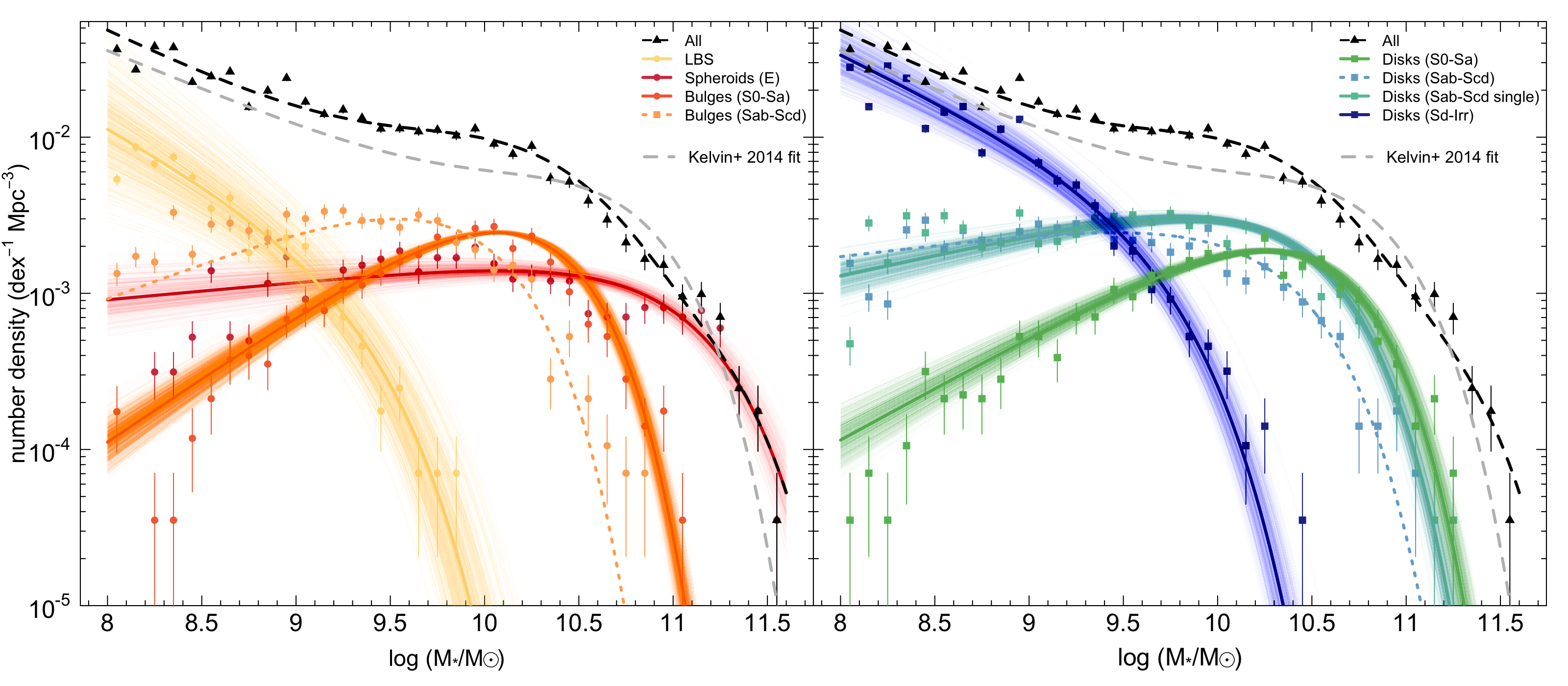}
\caption{Spheroid (left panel) and disk (right panel) stellar mass functions for different morphological classes, as fit by single Schechter functions. Although we do not fit directly to the binned galaxy counts, we show these data along with the fits for illustrative purposes, using common $1/V_{\rm max}$ weights for objects in 0.3 dex stellar mass bins as defined by \citet{Lange2015} and with Poisson error bars on the data counts. Error ranges for the individual MSMF fits are indicated by sampling 1000 times from the full posterior probability distribution of the fit parameters and plotting the resulting sampled mass functions with transparency such that darker regions indicate roughly one sigma uncertainties on the fits. The combined mass function of all components is shown in black, and we also plot the double Schechter total mass function of \citet{Kelvin_mfunc} for comparison.}
\label{fig:indivMFs}
\end{figure*}

\subsection{Maximum Likelihood Stellar Mass Function Fits}
\label{MLfits}

Consistent with the approach of \citet{vismorph}, we define a sliding volume-limited subsample of our data with mass limits that vary as a function of redshift. \citet{Lange2015} previously defined the appropriate mass limits as a function of redshift to create individual volume-limited samples of GAMA II that are at least 97.7\% complete and unbiased with respect to galaxy colour. We fit a smooth function to the same mass limits as a function of redshift, given by M$_{\rm lim} =$ 4.45 $+$ 207.2$z$ $-$ 3339$z^{2}$ $+$ 18981$z^{3}$, and require the sample we use for mass function fitting to have stellar mass greater than the appropriate mass limit evaluated at its redshift (see Fig.\ \ref{fig:samp}). As in \citet{vismorph}, we also exclude a small number of objects from our sample (26), whose automatedly derived photometric apertures have been flagged as erroneously large and had been assigned erroneously high stellar mass estimates. These objects are primarily in the Sd-Irr class, by far the most numerous class in our sample. The exclusion of these few objects is expected to cause minimal mass incompleteness due to their small fractional contribution to their respective classes.

To derive fits to the stellar mass distributions of the spheroid and disk populations of GAMA II, we employ a parametric maximum likelihood fitting method (e.g., \citealp{SandageML}; \citealp{EfML}), which is also used by \citet{vismorph} to derive morphologically defined stellar mass function fits. Our approach is similar to that described by \citet{Robotham2010}, where the probability density function (PDF) for each galaxy in mass space is represented by a single \citet{Schec76} type functional form:

\begin{multline} 
  \Phi(\log M)d \log M = ln(10)\times\phi^{*}10^{log(M/M^{*})(\alpha+1)}\\
  \times\exp(-10^{ \log (M/M^{*})})d \log M 
\end{multline}

where M$^{*}$ is the characteristic mass corresponding to the position of the ``knee'' in the mass function, while $\alpha$ and $\phi^{*}$ refer to the low-mass slope of the mass function and the normalization constant, respectively.

For this fitting method, the PDF that represents each galaxy must integrate to a total probability of one over the stellar mass range of detection. Since our sample is apparent magnitude limited, the relevant stellar mass interval for this integration varies as a function of redshift, and for each galaxy in our sample, the lower integration limit is set by the sample mass limit at its redshift, i.e., the sliding sample mass limit function described previously. For the individual structural \emph{components} of multi-component galaxies, applying a lower integration limit set by the systemic mass limit would lead to integration limits that do not necessarily encompass the measured component mass itself (depending on the component-to-total-mass ratio). As a result, we take the lower integration limits for components to be equal to the systemic mass reduced by the component-to-total-mass ratio of each component. To avoid biasing the mass function fits for the separate components, we must also consider whether or not individual component masses would fall below the overall sample mass limit if they were found in isolation. Thus, we omit galaxy components from our fits if they are below our overall mass fitting limit. We do not attempt to fit mass distributions below a global limit of log(M$_{*}$/M$_{\odot}$) $= 8$, below which we expect significant surface-brightness-based incompleteness in GAMA (see \citealp{Baldry2012} for further details). Through this variable mass limit approach, each galaxy or galaxy component's PDF is normalised to account for our redshift-dependent selection function, analogous to the application of $V/V_{\rm max}$ sample weights.

The galaxy PDFs are summed over the entire chosen sample to give the likelihood function that is then maximized to derive the most likely Schechter $\alpha$ and M$^{*}$ parameters. We use a Markov Chain Monte Carlo (MCMC) procedure for this analysis, implemented in the contributed \textsf{R} package $LaplacesDemon$\footnote{https://github.com/asgr/Laplacesdemon}. We choose to use the Componentwise Hit-And-Run Metropolis (CHARM) algorithm in this package and specify only a flat/uniform prior on fit parameters. We perform a minimum of 10,000 iterations for each fit (fits are also carried out 10 times for each class in order to derive jackknife errors on the fit parameters as discussed in \S \ref{res}) but also check for convergence using the $Consort$ function of $LaplacesDemon$\ and increase iterations performed for some classes where necessary. Since this procedure does not directly fit for the overall $\phi^{*}$ normalization parameter, we derive this value for each population through comparison to its observed number density. We require that the integrated Schechter function match the summed galaxy number distribution over a mass interval in which galaxy populations are well sampled (9 $<$ log(M$_{*}$/M$_{\odot}$) $<$ 10 for all types except Es where we sum up to log(M$_{*}$/M$_{\odot}$) $=$ 11 for improved statistics).

\section{Results}
\label{res}

\begin{figure}
\includegraphics[width=0.48\textwidth]{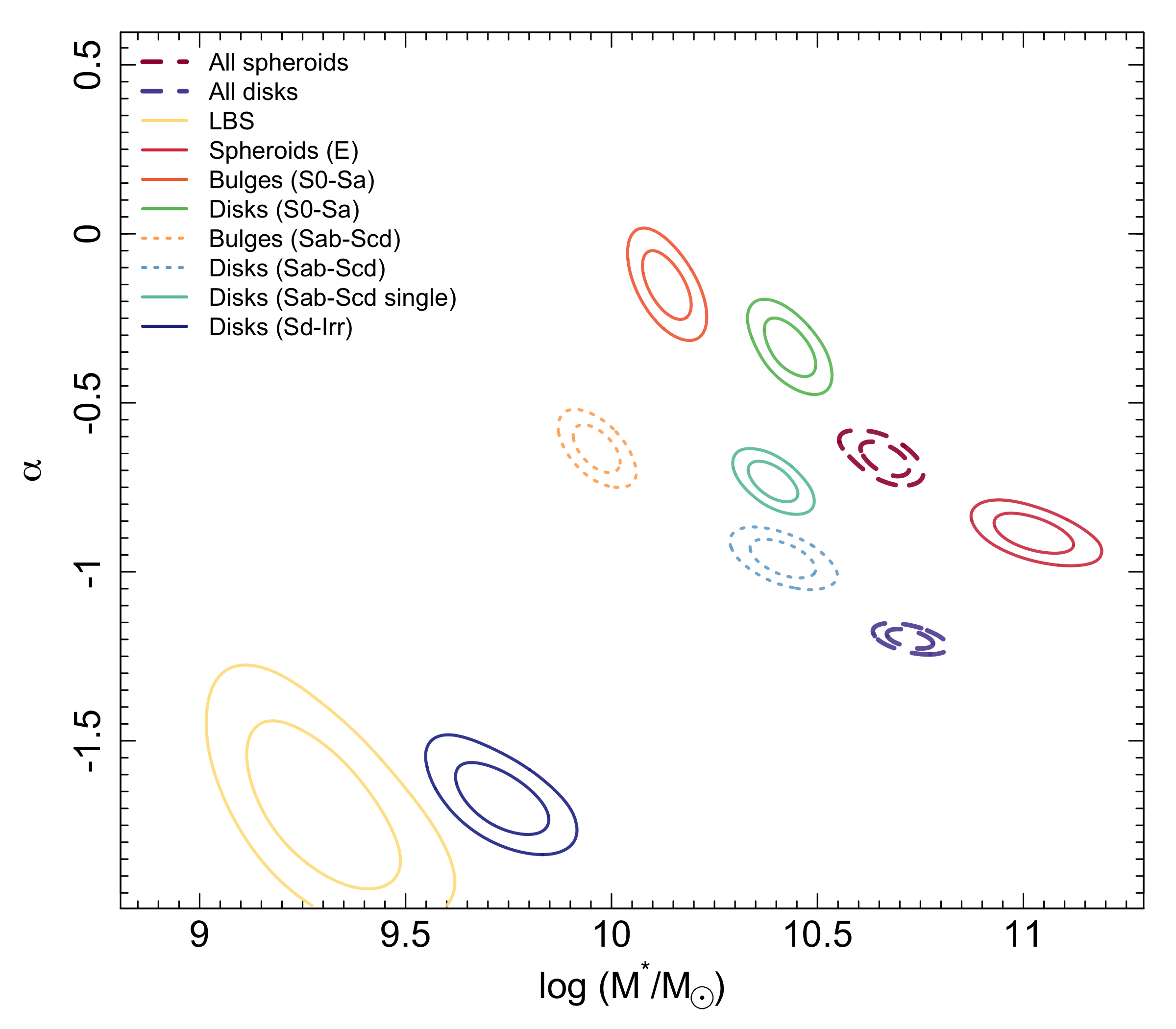}
\caption{One- and two-sigma error contours for separate spheroid and disk stellar mass function fits, divided by morphological type. Contours are derived from a jackknife resampling procedure that considers 10 subvolumes and the two-dimensional posterior probability distributions of all resulting fits.}
\label{fig:errcont}
\end{figure}

\begin{figure}
\includegraphics[width=0.48\textwidth]{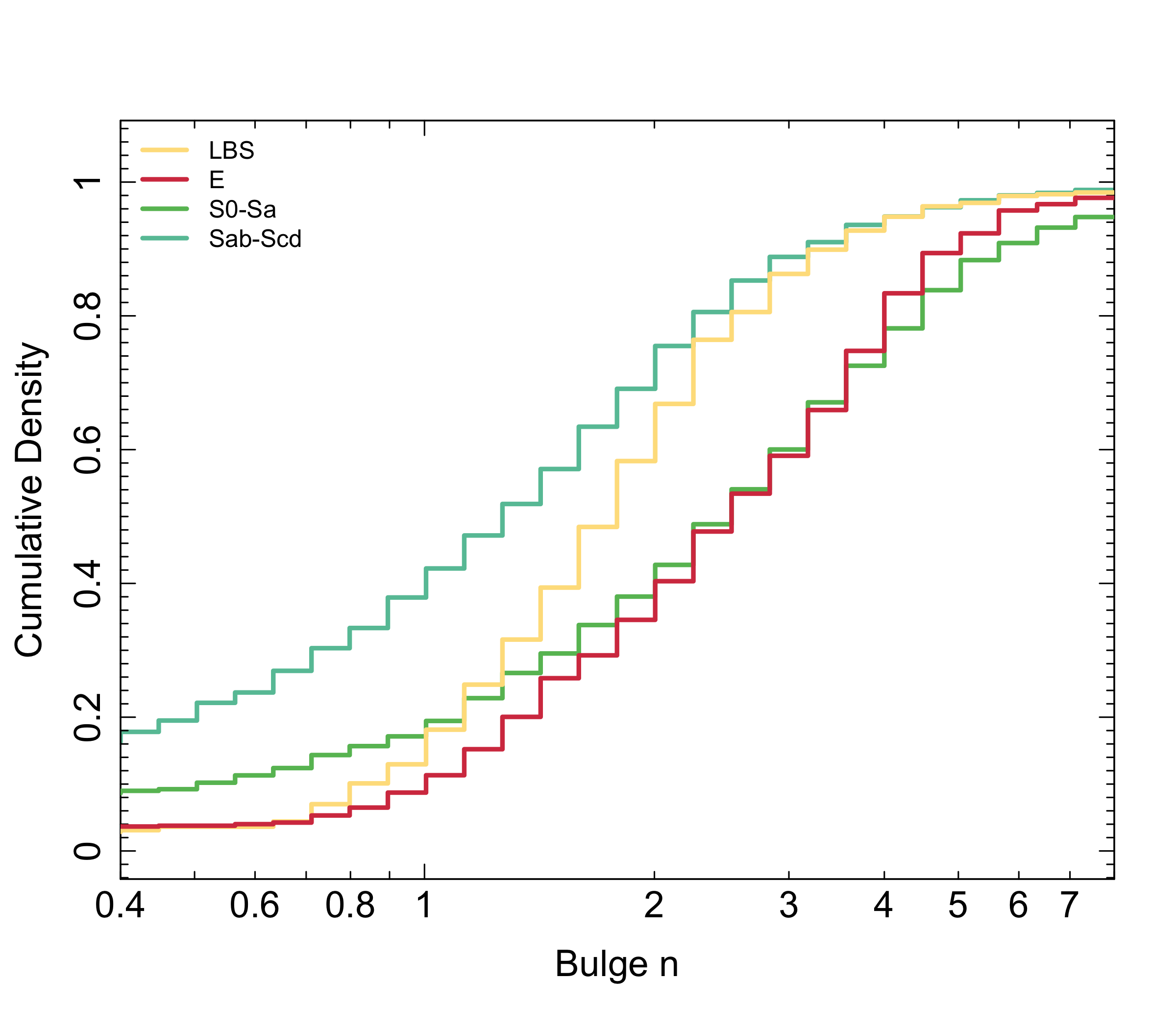}
\caption{Cumulative distribution of bulge S\'ersic index values by morphological type. The majority of Sab-Scd bulges display low S\'ersic indices more consistent with disky or pseudobulge structures than classical bulges.}
\label{fig:bulgen}
\end{figure}

Fig.\ \ref{fig:indivMFs} illustrates the derived spheroid and disk stellar mass function fits for the individual morphological type categories in our sample (fit parameters reported in Table \ref{tab:tab1} and binned mass function data points provided in an electronic table with columns described in Table \ref{tab:tab2new}). Single Schechter functions provide a reasonable description of each spheroid/disk population. For the morphological classes considered to be single-component systems (E, LBS, and Sd-Irr), these fits are effectively identical to the global morphological type stellar mass function fits reported by \citet{vismorph}, which expanded on the GAMA phase I analysis of \citet{Kelvin_mfunc}. For the assumed multi-component systems (S0-Sa and Sab-Scd), we derive separate bulge (or central component) and disk (outer component) stellar mass function fits. In both multi-component populations, the bulge and disk stellar mass functions differ significantly for the same galaxy type. Differences in M$^{*}$ and $\alpha$ Schechter-function parameters between separate populations are illustrated in Fig.\ \ref{fig:errcont} along with their associated error contours. To derive robust error contours, we use a jackknife resampling procedure that divides our sample into 10 subvolumes and consider the full two-dimensional posterior probability distributions for the parameters of all resulting fits. For the Sab-Scd population, we also illustrate the stellar mass function fit contours derived from a single-component treatment of Sab-Scds in addition to the individual component fits. As we motivate in the next section, we will choose to proceed with this single-component parameterisation of the Sab-Scd population when deriving total mass estimates.

\begin{figure*}
\includegraphics[width=0.49\textwidth]{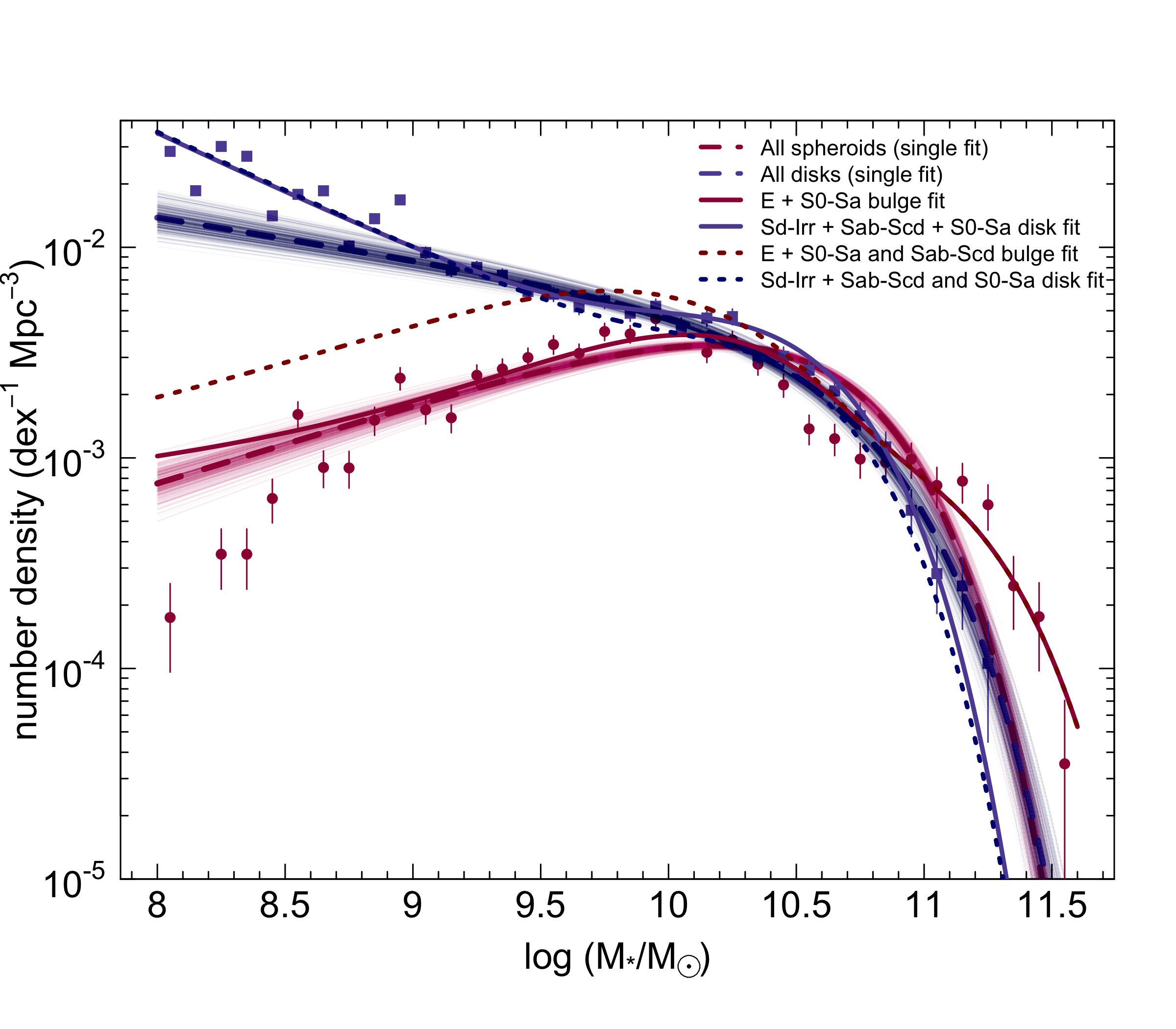}
\includegraphics[width=0.49\textwidth]{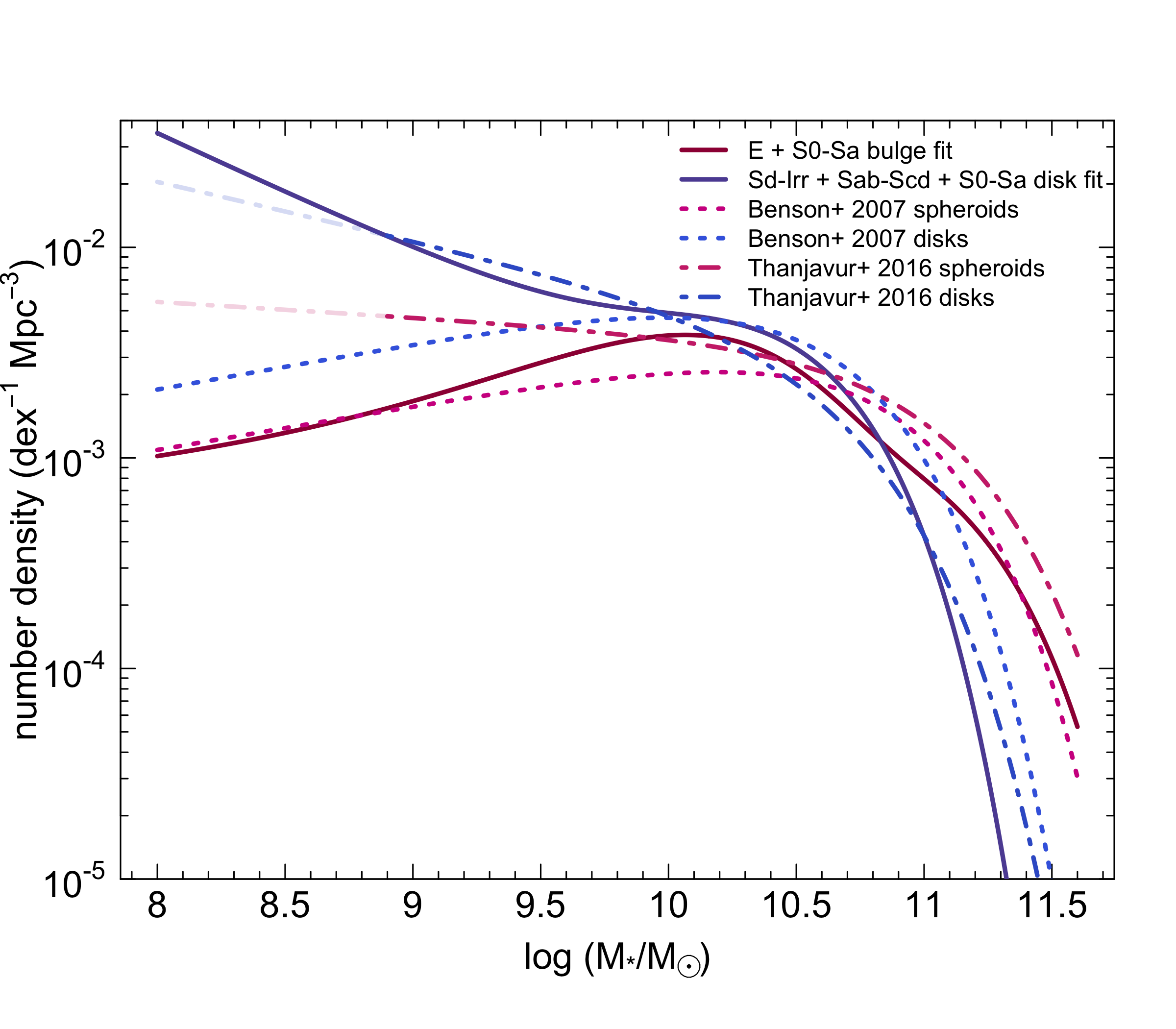}
\caption{Combined spheroid and disk stellar mass distributions. The left panel shows the spheroid and disk populations fit by single Schechter functions (dark red and blue points and dashed lines, respectively). However, these combined functions are better fit by the summed Schechter function fits to their individual constituents (solid red and blue lines). For comparison, we also show the combined spheroid and disk mass functions that would be derived if Sab-Scd central components were assigned to the spheroid class and Sab-Scd outer components were assigned to the disk class (dotted lines). The right panel compares our preferred combined spheroid and disk mass function fits to those of other authors. Spheroid and disk stellar mass functions from \citet{Benson2007} are plotted with an arbitrary normalisation for comparison purposes (red and blue dotted lines), and the equivalent mass functions from \citet{Thanjavur2016} are plotted as red and blue dot-dashed lines (light-coloured line segments indicate the extrapolation of these mass functions below the authors' mass limit). Data point weights and error ranges are indicated as in Fig.\ \ref{fig:indivMFs}.}
\label{fig:totMFs}
\end{figure*}

\subsection{Combined Spheroid and Disk Stellar Mass Functions}
\label{func_comp}
To construct combined mass functions for all spheroid-like and disk-like populations, we consider the single-component systems in the E category to consist of pure spheroids. We consider the single-component systems in the Sd-Irr category to consist of pure disks. As illustrated in Fig. \ref{fig:bulgen}, the LBS population displays a bulge S\'ersic n distribution that appears more skewed to low n values than the prototypical spheroids of the E population. This would seem to suggest that LBS galaxies may not closely resemble typical spheroids but rather have more in common with ``pseudobulges'' that typically display bulge n $\leq$ 2 (e.g., \citealp{KK2004}; \citealp{FD2008}). Pseudobulges are believed to differ from Es/classical bulges structurally, more closely resembling rotating disks (e.g., \citealp{Carollo1999}; \citealp{KK2004}). However, considerable ambiguity remains regarding the possibility of separating classical bulge and pseudobulge populations (see the review of \citealp{Graham2013} and references therein). Adding further complexity, \citet{Lange_decomp} find that the mass vs.\ size relation of LBSs is actually compatible with that of Es. Thus, with this ambiguity in mind, we refrain from including this population within either combined spheroid or disk mass function fit at this time and choose to report mass totals for this population separately.

For the multi-component systems of S0-Sa and Sab-Scd types, the obvious choice is to consider the central/bulge component of each class as a part of the spheroid population and the outer component as a part of the disk population. However, as shown in Fig. \ref{fig:bulgen}, the Sab-Scd bulge S\'ersic index (n) distribution again suggests typically low n values consistent with the pseudobulge population. In this case, the findings of \citet{Lange_decomp} also support the association of Sab-Scd bulges with the pure disk Sd-Irr populations. In addition, \citet{Lange_decomp} find that single S\'ersic fits are sufficient to describe the Sab-Scd population, yielding characteristics that are in similarly good agreement with the pure disk population. Thus, in subsequent fits, we elect to include the relation derived from single-component fits to the Sab-Scd population in the combined disk stellar mass function.

In Fig.\ \ref{fig:totMFs}, we show combined spheroid (E plus S0-Sa bulge) and disk (Sab-Scd, Sd-Irr, and S0-Sa disk) stellar mass distributions and Schechter function fits. Both combined spheroid and disk stellar mass functions are poorly fit by a single Schechter function form. As a result, we use the sum of the individual E and S0-Sa bulge Schechter function fits to describe the total spheroid mass distribution and the sum of Sab-Scd, Sd-Irr, and S0-Sa disk Schechter function fits to describe the total disk mass distribution. The low mass end of the combined spheroid mass distribution still deviates from this combined function slightly, which is largely due to deviations of the E mass function from the best-fitting Schechter function in the lowest few mass bins.

In the right panel of Fig.\ \ref{fig:totMFs}, we also compare to the prior combined spheroid disk and stellar mass function results of \citet{Benson2007} and \citet{Thanjavur2016}. Both sets of results were derived from bulge and disk decomposition analysis of SDSS imaging and were limited in depth by SDSS redshift survey sample magnitude limit ($>$2 mags brighter than our current sample). \citet{Thanjavur2016} specifically do not fit mass functions below log(M$_{*}$/M$_{\odot}$) $= 8.9$. We indicate the extrapolation of the \citet{Thanjavur2016} mass functions to our nearly one dex lower mass limit by the light-coloured line segments in Fig.\ \ref{fig:totMFs}.

Evidently, the \citet{Benson2007} spheroid mass function strongly resembles the spheroid mass function derived in this work, however, the combined disk mass functions diverge significantly, particularly at low mass. The low-mass disk mass function slope we derive is significantly steeper than that of \citet{Benson2007}, which suggests that this slope was not well constrained in the earlier, relatively shallow sample. The disk mass function of \citet{Thanjavur2016} follows the general shape of our disk mass function over the log(M$_{*}$/M$_{\odot}$) $> 8.9$ fitting region. However, the detailed shapes of these mass functions differ, as the \citet{Thanjavur2016} mass functions are parameterised as single Schechter functions in contrast to our multiple Schechter function combinations.

Our spheroid mass function differs significantly from the \citet{Thanjavur2016} mass function at both the high and low mass end. \citet{Bernardi2013} specifically discuss apparent discrepancies in the high-mass end of the mass function with reference to earlier GAMA-based and SDSS-based results. \citet{Bernardi2013} find that the same $z < 0.06$ upper redshift limit that we currently use eliminates the highest luminosity objects that overlap between the samples. As a result, it is possible that part of this disagreement originates from the smaller volume of GAMA, which implies poorer sampling of relatively rare high-mass galaxies. As discussed by \citet{Bernardi2013}, differences in the mass-to-light ratios assumed for high-mass galaxies can also cause such discrepancies. The reason for our discrepancy compared to the \citet{Thanjavur2016} spheroid mass function at low mass is less clear, however, it likely results from differences in the assignment of components to bulge and disk categories. \citet{Thanjavur2016} use a purely algorithmic approach to assigning galaxies to single or multi-component fit categories, which is based on cuts in the probability of various bulge plus disk or single S\'ersic models. We use the visual morphology as a prior on the single or multi-component status, and as a result of the morphology distribution of our sample, the majority of low-mass objects in our sample are fit as single-component, pure disk systems. While the typical B/T values derived by  \citet{Thanjavur2016} are low at low mass, these bulges added together create a spheroid mass distribution with a relatively flat low-mass slope. It is currently unclear whether these low-mass bulges are more consistent with disky pseudobulges or classical spheroids.

\begin{figure*}
\includegraphics[width=0.8\textwidth]{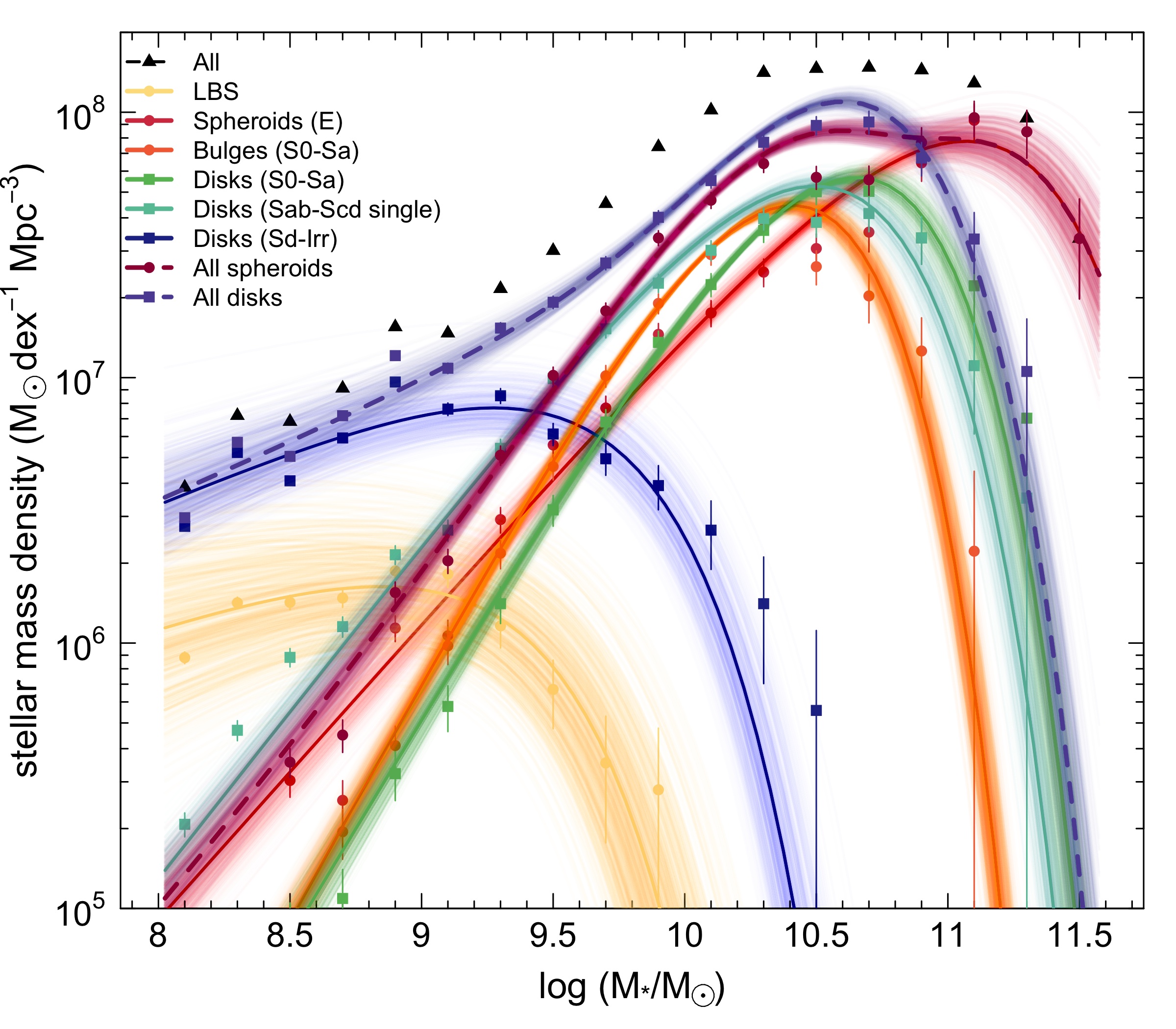}
\caption{Total mass density of spheroids and disks in separate classes, where points indicate the data values (with $1/V_{\rm max}$ weights), and lines indicate values derived from our Schechter function fits. Mass density estimates are bounded for each individual class. Error ranges on these fits are indicated as in Figs.\ \ref{fig:indivMFs} and \ref{fig:totMFs}.}
\label{fig:totmass}
\end{figure*}

\begin{figure*}
\includegraphics[width=0.9\textwidth]{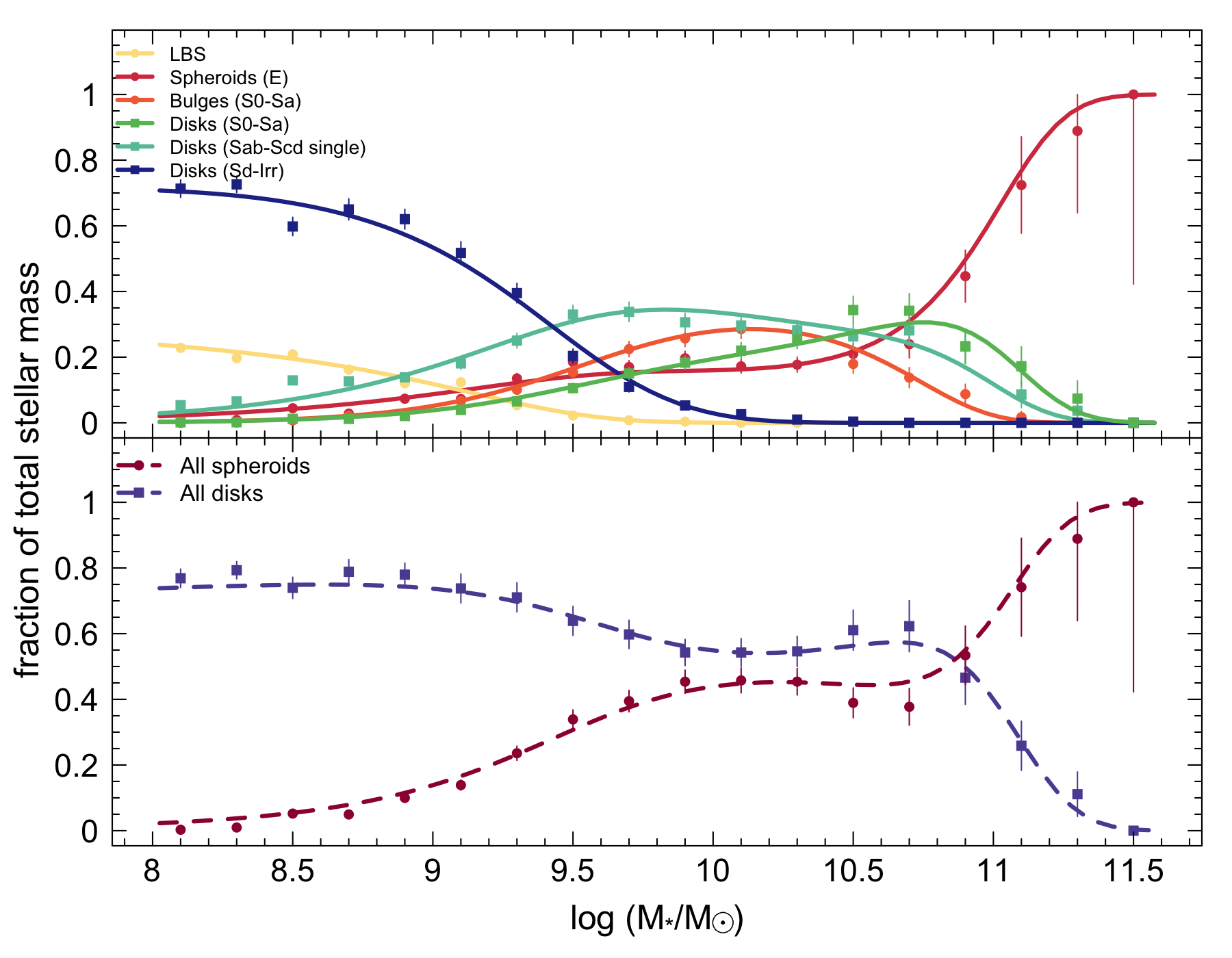}
\caption{Fraction of total stellar mass contributed by each population as a function of stellar mass regime, where points indicate data totals and lines are derived from our Schechter function fits. Due to their indeterminate nature, LBSs are excluded from combined spheroid and disk categories.}
\label{fig:massratio}
\end{figure*}

\subsection{Total Spheroid and Disk Mass Densities}
\label{mbreakdown}
Fig. \ref{fig:totmass} illustrates the total stellar mass density ($\rho_{*}$) values of spheroid/disk populations as a function of the stellar mass interval. For each spheroid/disk category, the peak of the stellar mass density distribution is well sampled, and our total stellar mass density estimates appear to be bounded within the limits of this sample. We derive total stellar mass density estimates for each structural category from both direct data summation using $V/V_{\rm max}$ weights ($\rho_{\Sigma}$) and integration of our stellar mass function fits ($\rho_{\phi}$). Table \ref{tab:tab2} summarizes the stellar mass density estimates along with uncertainties derived using the same jackknife resampling procedure as in \citet{vismorph}. All such estimates are subject to an additional error term from cosmic variance. With the method of \citet{DR2010}, we estimate a 22.3\% cosmic variance error contribution within our sample volume.

Integrating our combined stellar mass function fits, we find a total spheroid stellar mass density $\rho_{spheroid} = 1.24\pm 0.49 \times 10^{8}$ M$_{\odot}$Mpc$^{-3}$h$_{0.7}$, which translates to $\sim$50\% of the total stellar mass density. Breaking down the mass density further, 35\% of the total is contributed by Es, and 15\% is contributed by S0-Sa bulges. Disk-like structures are found to have mass density $\rho_{disk} = 1.20\pm 0.45 \times 10^{8}$ M$_{\odot}$Mpc$^{-3}$h$_{0.7}$, which translates to a similar $\sim$48\% of the total. The disk population contributions to the total are 22\% in Sab-Scd galaxies, 6\% in Sd-Irr galaxies, and 20\% in S0-Sa disks. The remaining few percent of the total stellar mass density is found in the ambiguous LBS class.

The spheroid and disk mass ratios we derive are broadly consistent with previous results, including the approximately equal spheroid/disk mass ratio estimated by \citet{Kelvin_mfunc} and \citet{vismorph}. Bracketing our result, \citet{Benson2007} estimated a disk mass fraction of 35-51\%, where the lower fraction is determined with a correction to the \emph{luminosity} function bias in the sample inclination distribution (see e.g., \citealp{TW2011}). Similarly, \citet{Gadotti2009} estimated a lower 36\% disk mass fraction but in a sample with a mass limit 2 dex higher than the current work. \citet{Gadotti2009} also discuss the comparison to samples with lower mass limit and find that their spheroid/disk mass fractions would indeed be approximately equal within a sample with a significantly lower mass limit.

Similarly, the recent work of \citet{Thanjavur2016} estimates a 37\% disk mass fraction in a sample with a mass limit approximately one dex higher than the current work. As discussed in \S \ref{func_comp}, the higher spheroid mass fraction results from discrepancies with our spheroid mass function at both high and low masses. It is interesting to note that our total mass fraction discrepancy with this result could potentially be resolved through treating our Sab-Scd galaxies as two-component systems. Assuming that the central components of these systems add to the spheroid mass and the outer components add to the disk mass is likely more similar to the \citet{Thanjavur2016} component treatment. In this case, we would find a total disk mass fraction of 39\% and a spheroid mass fraction of 59\%. However, we find that our actual spheroid mass \emph{function} in this case would still deviate significantly from the \citet{Thanjavur2016} spheroid mass function, as this change primarily affects the shape of the mass function at intermediate masses rather than at low or high mass (see dotted lines in the left panel of Fig.\ \ref{fig:totMFs}). 

Compared both to the current work and to other authors, \citet{Driver2007full} derive a slightly higher disk mass fraction of 59\%. The higher disk mass fraction may be due in part to the deeper-than-SDSS imaging used in the \citet{Driver2007full} analysis, which should enable detection of the outskirts of galaxy disks to lower surface brightness levels than we are able to reach here. Further, the \citet{Driver2007full} analysis uses the Millennium Galaxy Catalogue (MGC; \citealp{MGC}) sample, which is $B$-band selected and may plausibly include a larger fraction of blue and likely disk-like objects at fixed \emph{mass} than our $r$-band selected sample.

The measured balance of spheroid and disk stellar mass at z$\sim$0 provides a fundamental constraint on galaxy formation and evolution models, as it effectively results from the detailed interplay between structure formation and destruction processes as they build up the galaxy population over cosmic time. Although we find estimated disk stellar mass densities slightly lower than \citet{Driver2007full}, our spheroid and disk stellar mass densities are plausibly consistent with the predicted spheroid/disk stellar mass buildup from the two-phase galaxy formation model of \citet{Driver2013}, given the uncertainties and assumptions involved in both. Further, our mass density estimates agree well with an updated version of this model as presented by \citet{Andrewsmodel}.

\begin{figure*}
\includegraphics[width=0.49\textwidth]{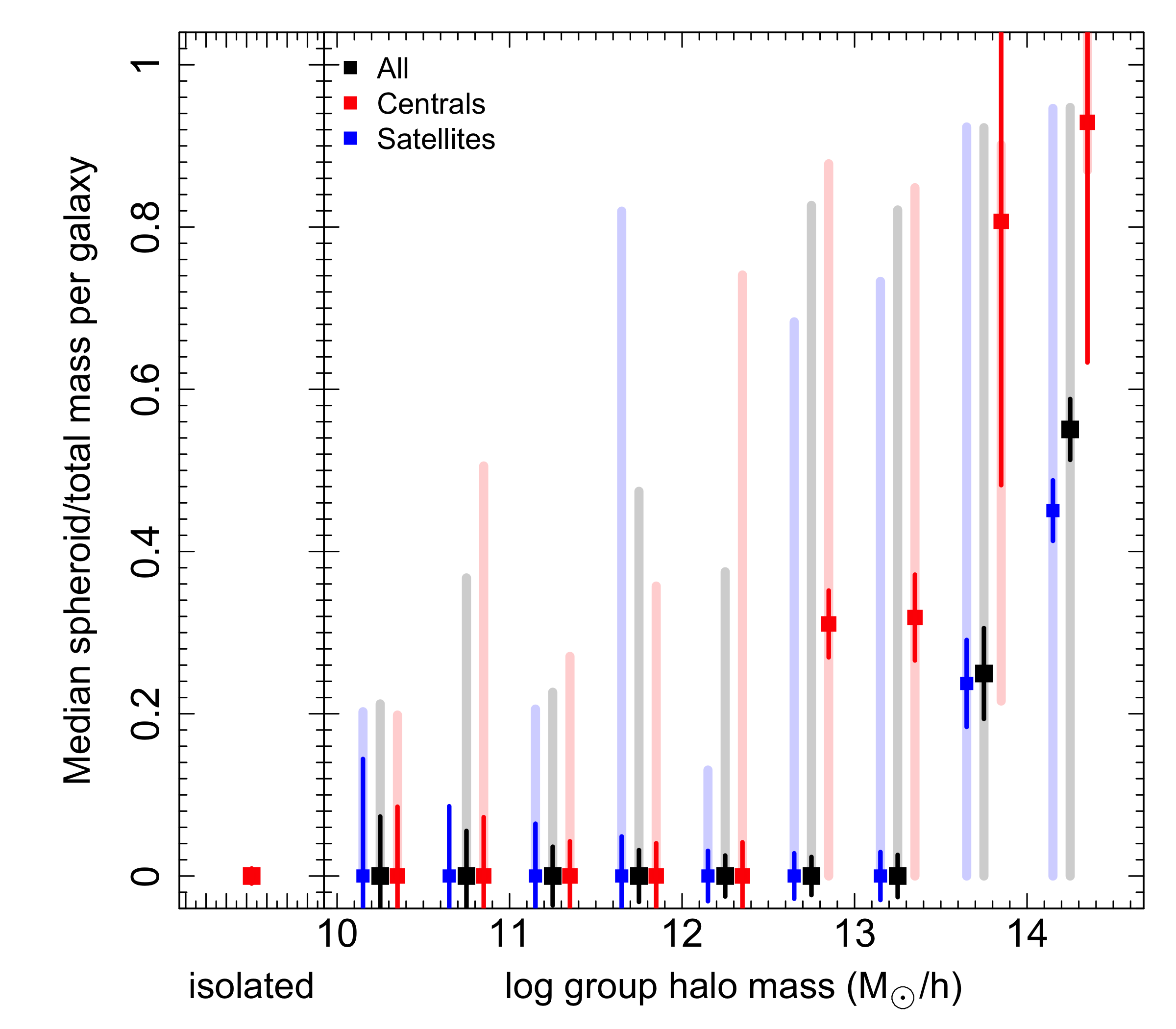}
\includegraphics[width=0.49\textwidth]{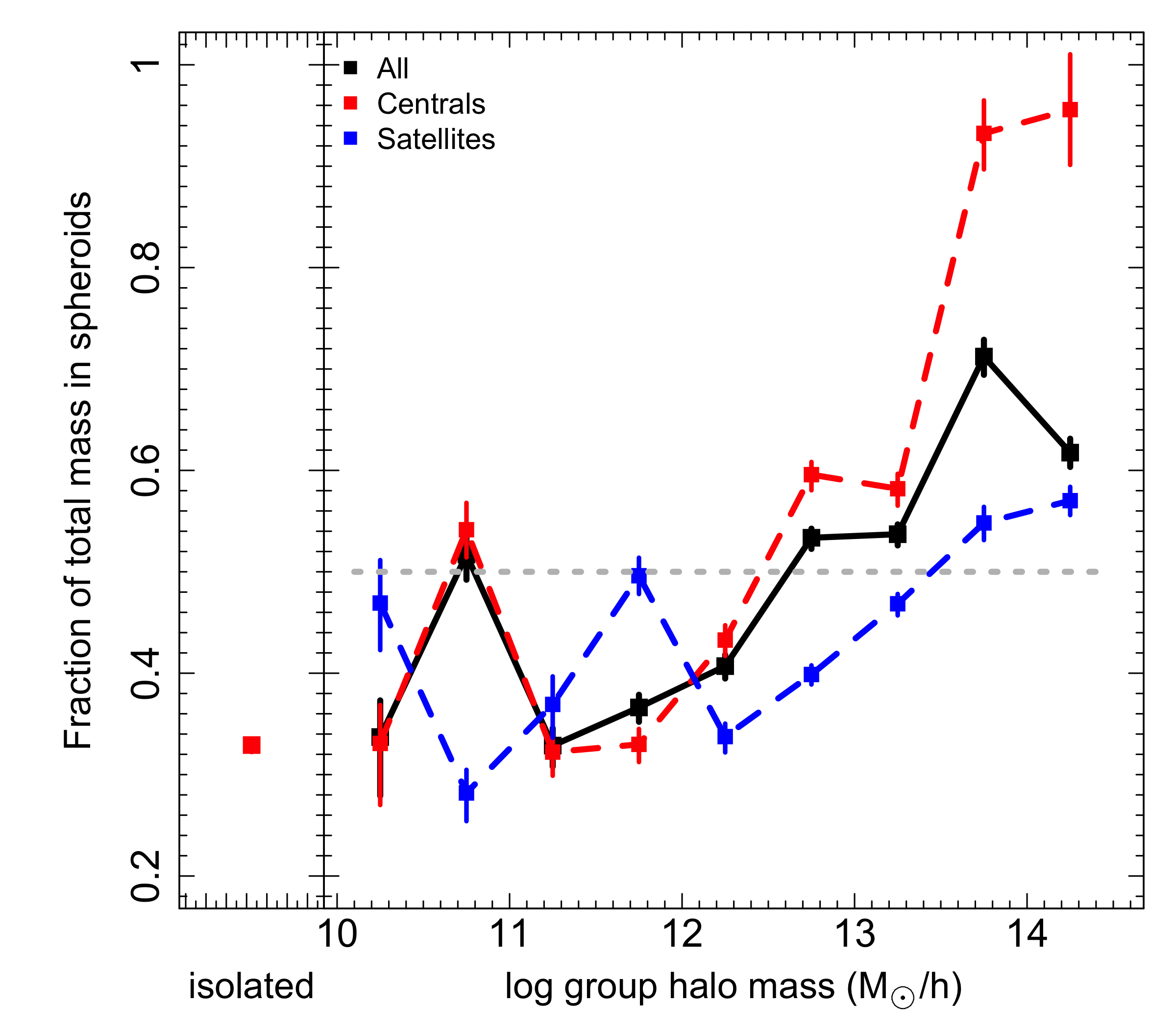}
\caption{Spheroid to total mass ratios as a function of group environment for central galaxies, satellite galaxies, and the combined population. The left panel shows the per-galaxy median (squares) with estimated one-sigma errors on the median (dark bars) and the interquartile range of the data (light bars), with central and satellite points shown offset from the bin centers for clarity. The right panel shows the summed total for all objects in each bin with one-sigma error bars on the fraction in each bin indicated by vertical bars. Due to their indeterminate nature, LBSs are excluded from either spheroid or disk category here. For points with no apparent vertical bars, the one-sigma errors and/or interquartile ranges are smaller than the points.}
\label{fig:env}
\end{figure*}

\subsection{Variation of the Spheroid and Disk Stellar Mass Budget}

Aside from the global mass balance, the detailed balance between galaxy spheroid and disk mass buildup to z$\sim$0 \emph{as a function of galaxy mass and environment} can be measured in both observations and galaxy evolution models. In the following section, we quantify such variations in the spheroid and disk mass budget using GAMA survey observations.

\subsubsection{Spheroid and Disk Mass as a Function of Galaxy Mass}
In Fig. \ref{fig:massratio}, we show the fraction that each spheroid/disk category in our sample contributes to the total stellar mass density in each galaxy mass bin. The trends shown in this figure are complex, but they reflect a number of expected large-scale galaxy demographic trends, such as the transition from spheroid mass dominance at high mass to disk mass dominance at low mass. For individual galaxy types, we see the E mass dominance at the highest stellar masses give way to S0-Sa disks and bulges at lower mass, then to Sab-Scd galaxies with a broad distribution through the intermediate mass regime, and finally to dwarf Sd-Irr disks with a smaller contribution from LBSs at the lowest masses we probe. The transition between overall spheroid and disk mass dominance occurs at log(M$_{*}$/M$_{\odot}$) $\sim 10.9$ just above the bimodality mass of \citet{bimodalitymass} at log(M$_{*}$/M$_{\odot}$) $\sim 10.5$, where quenched (and presumably spheroid-dominated) galaxies give way to those with recent star formation. \citet{Thanjavur2016} show qualitatively similar trends in the spheroid and disk mass ratios in their Fig.\ 11, but we find more detailed structure in the trends with mass compared to the smooth variation seen in the other work. \citet{Thanjavur2016} also find the transition point between spheroid and disk mass dominance occurs at a slightly lower mass than we find, closer to the bimodality mass.

\begin{figure*}
\includegraphics[width=0.49\textwidth]{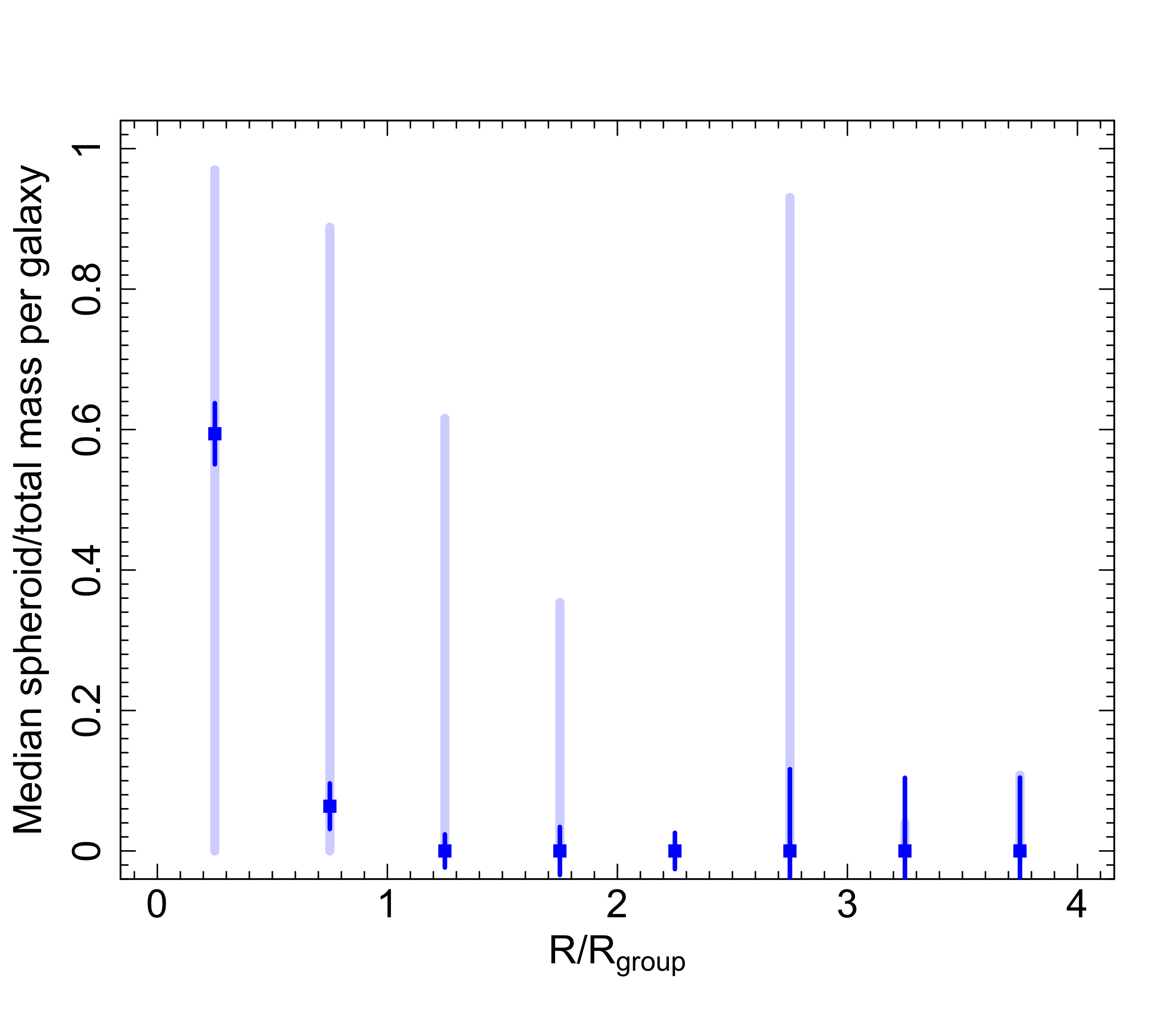}
\includegraphics[width=0.49\textwidth]{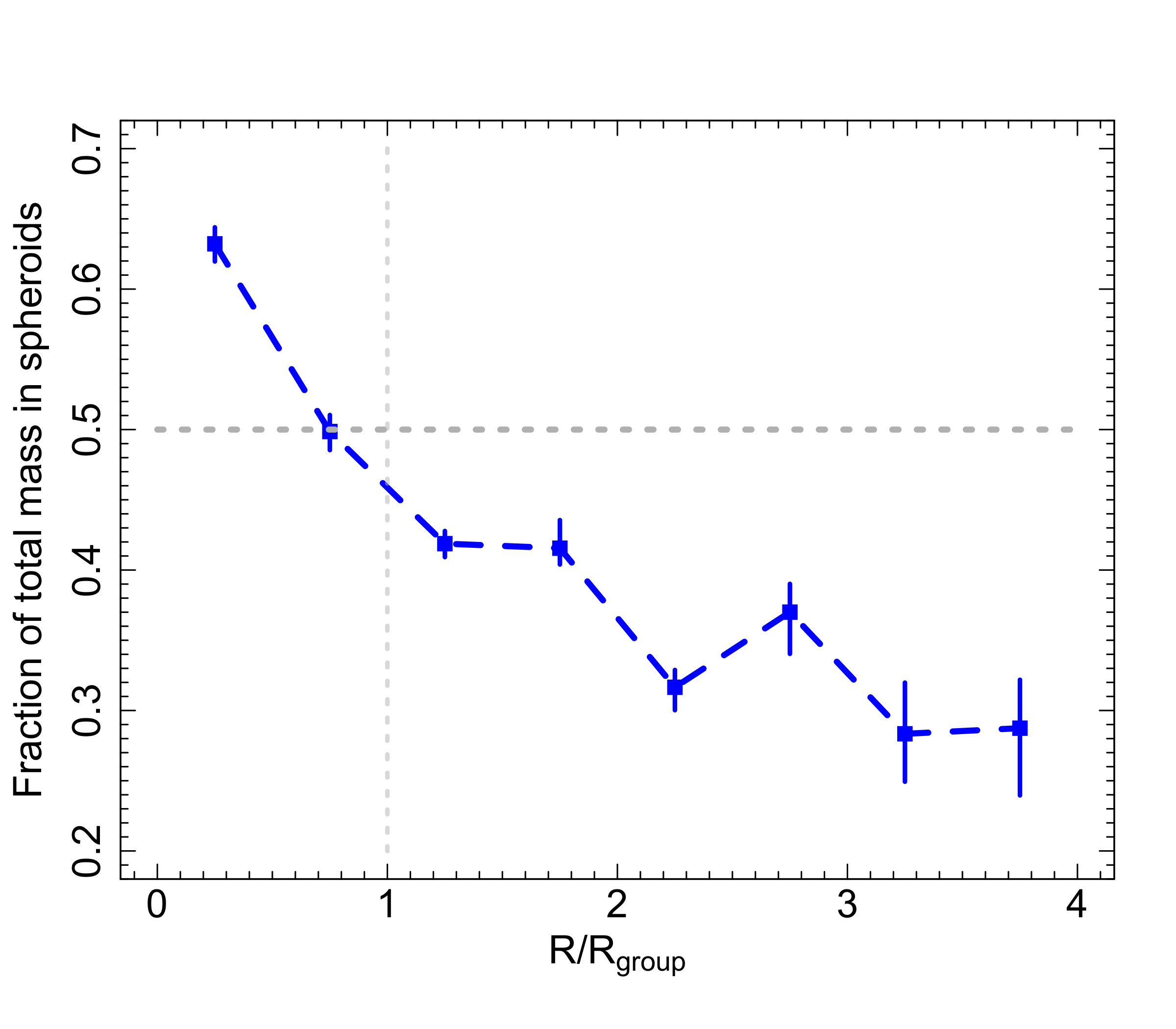}
\caption{Spheroid to total mass ratios for satellite galaxies, as a function of projected radius from the group center position. The left panel shows the per-galaxy median (squares) with estimated one-sigma errors on the median (dark bars) and the interquartile range of the data (light bars). The right panel shows the summed total for all objects in each bin with one-sigma error bars on the fraction in each bin indicated by vertical bars (due to their indeterminate nature, LBSs are excluded from either spheroid or disk category here).}
\label{fig:envrad}
\end{figure*}

\subsubsection{Spheroid and Disk Mass as a Function of Environment}
    Galaxy structure is well known to vary with the surrounding environment, as through the ``morphology-density relation'' (e.g., \citealp{Dressler1980}). In Fig.\ \ref{fig:env} we examine the balance between spheroid and disk mass as a function of group halo environment specifically, using group identifications derived in the GAMA II group catalog of \citet{GAMAgroups}. We also show the division between group central and satellite galaxies as derived from this catalog. We find that $\sim$54\% of the present sample are considered isolated in the GAMA group-finding analysis, i.e., in N$=$1 halos. We indicate the spheroid-to-total ratios for these points by the red ``isolated'' points in this figure. A small number ($<$40) of our sample galaxies are found in slightly lower mass groups than we plot here. However, any bins with M$_{halo} < 10^{10} M_{\odot}/h$ are sparsely populated and dominated by low-N groups (N$\leq$4) for which derived group halo mass estimates are less reliable, so we refrain from analysing these lower mass systems here.

 The left panel of this figure illustrates the median and spread in the distribution of spheroid mass divided by total mass of individual galaxies in both isolated and grouped environments. From this figure, we find that the isolated objects are primarily disk dominated, whle there is extremely large spread in individual galaxy spheroid-to-total-mass ratios within each halo mass bin. In general, similar degrees of spread in spheroid-to-total-mass ratios are found for both central and satellite galaxies, which could indicate that this spread is driven in part by group-to-group variations within each halo mass bin. The median trend for satellite galaxies (and for the combined sample) rapidly flattens to a typically zero spheroid mass ratio (i.e., pure disk) by group halo mass $\sim10^{13} M_{\odot}/h$, however, the typical spheroid mass fraction among central galaxies remains nonzero to slightly lower group mass $\sim10^{12.5} M_{\odot}/h$. This marginal difference is a likely consequence of the previously discussed correlation between galaxy mass and spheroid mass ratio, as centrals tend to be more massive than satellites within a given halo mass bin and are correspondingly more likely to be spheroid dominated.

 The right panel of Fig.\ \ref{fig:env} examines the relationship between spheroid mass ratio and group environment in an integrated sense, where we sum the total stellar mass of all objects in each group environment bin and plot the total spheroid mass in each bin divided by the total stellar mass of all components. We find a strong decrease in the spheroid mass fraction going from high to low group halo masses, with the mass fraction for low mass groups similar to that for isolated systems. For satellite galaxies, spheroids only dominate the mass budget for the highest mass groups we probe, above M$_{halo} \sim 10^{13.5} M_{\odot}/h$. The transition between integrated spheroid and disk mass dominance for group central galaxies occurs at a lower group halo mass M$_{halo} \sim 10^{12.5} M_{\odot}/h$, again likely reflecting the positive correlation between galaxy mass and spheroid mass ratio. 

The simulations of \citet{Sales2012} examined the role of group halo properties in galaxy spheroid/disk formation and found galaxy structure to be poorly correlated with host halo properties but strongly correlated with the alignment of gas accreted into the halo. However, only a narrow range of galaxy host halo masses were considered for this analysis (similar to the halo mass of the Milky Way), which is within the regime where we find flat spheroid mass ratios as a function of group halo mass. In general, there is reason to expect a correlation between host halo environment and structure formation in simulations as well as observations. Halos in high density environments may be expected to collapse earlier than those in less dense environments and thus be more concentrated and likely to host lower angular momentum, more spheroid-dominated galaxies (e.g., as discussed by \citealp{RF2012}). Reproduction of the mass ratios of galaxy spheroid and disk structures observed across a variety of environments should provide a useful test of future developments in cosmological galaxy formation models.

With the GAMA dataset, we can also examine spheroid and disk mass trends internal to groups. For the satellite galaxy population specifically, it is likely that the spread in spheroid-to-total mass ratio at fixed halo mass is at least partially driven by residual variations of spheroid mass ratio with distance from each group's center. To investigate this trend, we use the projected distance of each galaxy from the iterative group center position (R) and scale these radii by a characteristic radius for each group, R$_{\rm group}$, which we take as the radius encompassing 50\% of the group members from \citet{GAMAgroups}.

Similar to Fig.\ \ref{fig:env}, Fig.\ \ref{fig:envrad} illustrates the variation of spheroid mass ratio with distance from the group center for both individual galaxies (left panel) and for summed totals in radius bins (right panel). We find that satellite galaxies still display significant per-galaxy variation in spheroid-to-total ratio at fixed radius, which implies that other factors such as group-to-group variations or galaxy mass segregation drive additional scatter at fixed radius. We note, however, that the recent analysis of \citet{Kafle_mseg} has found no evidence for mass segregation of the satellite population as a function of radius in the GAMA groups. In spite of the scatter at fixed radius, a clear trend exists whereby spheroid mass fraction increases as distance from the cluster center decreases. In the cluster outskirts, the typical satellite galaxy is disk dominated, and spheroid-dominated satellite galaxies are only the norm in the lowest radius bin we probe. In an integrated sense, the total mass budget for group satellites becomes spheroid dominated just below the characteristic 50th percentile group radius (as indicated by the dashed lines in Fig.\ \ref{fig:envrad}).

Qualitatively our observed mass ratio trend with radius matches the expectation from previous works where bulge-dominated \emph{morphology} is found to become more common in high density environments nearer to group/cluster cores (e.g., \citealp{Dressler1980}; \citealp{PG1984}; \citealp{Tran2001}; \citealp{Hoyle2012}). By casting this trend in terms of purely quantitative mass ratios, we intend to directly probe the regions of group parameter space in which spheroid and disk mass assembly processes dominate. We note that our error bars on these quantitative mass ratios can be large in certain regimes where sample numbers are low, particularly for centrals at large group halo mass and for the lower number density outskirts of groups. Future efforts to extend GAMA structural analysis outwards in redshift using higher resolution imaging should improve these constraints with a larger sample volume, but such constraints will necessarily apply over a larger redshift range than the z$\sim$0 results presented here.

\citet{LG2013} take a similar quantitative approach in measuring the disk-to-total mass ratios of galaxies, finding a very weak dependence of D/T on local projected fifth nearest neighbor density but a stronger trend between D/T and group crossing time (proportional to distance from the group center) that matches the sense of our mass ratio trend. \citet{LG2013} propose that galaxy harassment \citep{Moore1996}, which is most effective in high density regions where high-speed galaxy-galaxy encounters are likely, is a plausible explanation for this trend. In this scenario, our results would imply that galaxy harassment is most effective at converting disk mass to bulge mass in relatively rich group/cluster environments and within the 50th percentile group radius.

\begin{table*}
  \caption{\label{tab:tab1} Single Schechter stellar mass function fit parameters for the spheroid and disk stellar mass functions in Figs.~\ref{fig:indivMFs} and \ref{fig:totMFs}. Columns are: the knee in the Schechter function (M$^{*}$), the slope ($\alpha$), and the normalization constant ($\phi^{*}$). Quoted uncertainties are derived from the spread in each parameter's posterior probability distribution from fits carried out in 10 jackknife resampling iterations.}
\begin{tabular}{cccc} \hline
Population & log(M$^{*}$h$_{0.7}$$^{2}$/M$_{\odot}$) & $\alpha$ & $\phi^{*}/10^{-3}$  \\
&                  &          & (dex$^{-1}$Mpc$^{-3}$h$_{0.7}$$^{3}$) \\ \hline
E &$11.02 \pm 0.055$ & $-0.887 \pm 0.034$ & $0.866^{+0.080}_{-0.078}$ \\
S0-Sa bulges&$10.15 \pm 0.033$ & $-0.179 \pm 0.056$ & $2.84^{+0.089}_{-0.11}$ \\
S0-Sa disks&$10.43 \pm 0.036$ & $-0.337 \pm 0.050$ & $2.06^{+0.11}_{-0.12}$ \\
Sab-Scd bulges &$9.868 \pm 0.033$ & $-0.54 \pm 0.040$ & $2.94^{+0.11}_{-0.12}$ \\
Sab-Scd disks &$10.29 \pm 0.045$ & $-0.852 \pm 0.032$ & $1.63^{+0.10}_{-0.10}$ \\
Sab-Scd combined &$10.40 \pm 0.034$ & $-0.736 \pm 0.034$ & $2.42^{+0.15}_{-0.15}$ \\
Sd-Irr &$9.647 \pm 0.065$ & $-1.58 \pm 0.062$ & $1.67^{+0.42}_{-0.31}$ \\
LBS &$9.31 \pm 0.11$ & $-1.66 \pm 0.15$ & $0.713^{+0.37}_{-0.25}$ \\
 \hline
All spheroids &$10.60 \pm 0.035$ & $-0.623 \pm 0.029$ & $3.70^{+0.15}_{-0.15}$ \\
All disks &$10.73 \pm 0.033$ & $-1.20 \pm 0.016$ & $1.72^{+0.12}_{-0.12}$ \\
\hline
\end{tabular}
\end{table*}


\begin{table*}
  \caption{\label{tab:tab2new} Binned stellar mass function data points for individual galaxy populations, as shown in Figs.~\ref{fig:indivMFs} and  \ref{fig:totMFs}. This table is provided online in machine readable form, with columns as described below.}
\begin{tabular}{cc} \hline
Column number & Column description \\ \hline
1 & stellar mass bin midpoints \\
2-4 & E stellar mass function (lower bound, measurement, upper bound) \\
5-7 & S0-Sa bulge stellar mass function (lower bound, measurement, upper bound) \\
8-10 & S0-Sa disk stellar mass function (lower bound, measurement, upper bound) \\
11-13 & Sab-Scd bulge stellar mass function (lower bound, measurement, upper bound) \\
14-16 & Sab-Scd disk stellar mass function (lower bound, measurement, upper bound) \\
17-19 & Sab-Scd combined stellar mass function (lower bound, measurement, upper bound) \\
20-22 & Sd-Irr stellar mass function (lower bound, measurement, upper bound) \\
23-25 & LBS stellar mass function (lower bound, measurement, upper bound) \\
26-28 & All spheroid stellar mass function (lower bound, measurement, upper bound) \\
29-31 & All disk stellar mass function (lower bound, measurement, upper bound) \\
\hline
\end{tabular}
\end{table*}

\begin{table*}
 \caption{\label{tab:tab2} Stellar mass densities for each spheroid/disk category, derived both by summation of data with $V/V_{\rm max}$ weights ($\rho_{\Sigma}$) and integration of stellar mass functions ($\rho_{\phi}$). A fraction of the total stellar mass is also given for each category and method. Quoted uncertainties are derived according to a jackknife resampling procedure as decribed in \S \ref{mbreakdown}. Derived stellar mass density estimates are also subject to an additional 22.3\% error contribution from cosmic variance, estimated by the method of \citet{DR2010}.}
\begin{tabular}{ccccc} \hline
Population & $\rho_{\Sigma}/10^7$ & Fraction of All (sum)& $\rho_{\phi}/10^7$ & Fraction of All (fit)\\
           & (M$_{\odot}$Mpc$^{-3}$h$_{0.7}$) &              & (M$_{\odot}$Mpc$^{-3}$h$_{0.7}$) &  \\ \hline
All &$23 \pm 7.7$ & ... & $25 \pm 4.9$ & ... \\
 \hline
E &$8.3 \pm 2.9$ & $0.36$ & $8.6 \pm 2.1$ & $0.35$ \\
S0-Sa bulges&$3.5 \pm 1.1$ & $0.15$ & $3.8 \pm 1.4$ & $0.15$ \\
S0-Sa disks &$4.9 \pm 1.7$ & $0.21$ & $5.0 \pm 1.8$ & $0.20$ \\
Sab-Scd &$5.1 \pm 1.6$ & $0.22$ & $5.4 \pm 1.8$ & $0.22$ \\
Sd-Irr &$1.3 \pm 0.40$ & $0.054$ & $1.6 \pm 0.39$ & $0.063$ \\
LBS &$0.23 \pm 0.071$ & $0.0097$ & $0.37 \pm 0.20$ & $0.015$ \\
 \hline
All Spheroids &$12 \pm 4.0$ & $0.51$ & $12 \pm 4.9$ & $0.50$ \\
All Disks &$11 \pm 3.7$ & $0.48$ & $12 \pm 4.5$ & $0.48$ \\
\hline
\end{tabular}
\end{table*}

\section{Summary and Conclusions}
\label{conc}

Using the recently expanded Galaxy and Mass Assembly (GAMA) survey phase II visual morphology sample and the large-scale bulge and disk decomposition analysis of \citet{Lange_decomp}, we derive new stellar mass function fits to galaxy spheroid and disk populations down to log(M$_{*}$/M$_{\odot}$) $= 8$ . We find an approximately equal division between the total stellar mass densities of galaxy spheroid and disk populations, which is broadly consistent with prior results albeit with a somewhat lower disk mass fraction than observed by \citet{Driver2007full}. The fact that \citet{Driver2007full} used deeper imaging data than in our current analysis raises the intriguing possibility that the planned future extension of GAMA structural analysis to use deeper and higher-resolution Kilo-Degree Survey imaging (KiDS; \citealp{kids}) could yield disk galaxies undetected within our current surface brightness limits or larger and more massive disks in existing galaxies. The resolution of KiDS imaging will also allow us to extend the GAMA structural analysis to higher redshift, improving the sampling of high mass galaxies and potentially resolving a discrepancy with the high-mass end of the mass function as seen in the larger SDSS volume. Further, we find a small (few percent) of our total stellar mass density in the LBS (little blue spheroid) population, which is not a clearly identified as either a spheroid or disk population at present. Future investigations with KiDS imaging should allow us to better resolve the structural characteristics of these objects, including their potential for hosting low-surface-brightness outer envelopes.

Finally, we examine the variation of the total disk and spheroid mass balance as a function of galaxy mass and group environment. We find strong overall population trends with both galaxy mass and group halo mass, where spheroids dominate the galaxy mass budget above galaxy stellar mass $\sim10^{11}$ M$_{\odot}$ and above group halo mass $\sim10^{12.5}$ M$_{\odot}/h$. Further, we find differences in the mass budget of satellites and centrals, where satellites are only spheroid dominated within higher group halo mass environments (M$_{halo} > 10^{13.5} M_{\odot}/h$). This difference is related to the typically lower masses of satellites compared to centrals at fixed halo mass. We also examine satellite galaxy spheroid-to-total mass ratio trends with radius from the group center, finding that spheroids dominate the mass budget of satellite galaxies within the 50th percentile group radius. This trend towards spheroid dominance at low group-centric radius is likely due to mechanisms that are most effective at transforming morphology where galaxy densities and encounter speeds are high, such as galaxy harassment \citep{Moore1996}.

    These measurements, which are currently possible from photometric galaxy decompositions in large survey samples, provide a useful basis for comparison with the detailed structural demographics of simulated galaxies. In the future as samples of kinematic galaxy surveys continue to grow (e.g., from the SAMI survey of ~3,000 galaxies to the MaNGA survey of ~10,000 galaxies; \citealp{SAMI}; \citealp{manga}), the division between spheroid-like and disk-like galaxy \emph{dynamics} will be possible on similarly large scales, providing an even more direct constraint on models of galaxy structural evolution.

\section*{Acknowledgements}

We thank Alister Graham, Anne Samson, and the anonymous referee for helpful comments on this manuscript. SPD and AJM acknowledge funding support from the Australian Research Council under Discovery Project 130103505.

GAMA is a joint European-Australasian project based around a spectroscopic campaign using the Anglo-Australian Telescope. The GAMA input catalogue is based on data taken from the Sloan Digital Sky Survey and the UKIRT Infrared Deep Sky Survey. Complementary imaging of the GAMA regions is being obtained by a number of independent survey programmes including GALEX MIS, VST KiDS, VISTA VIKING, WISE, Herschel-ATLAS, GMRT and ASKAP providing UV to radio coverage. GAMA is funded by the STFC (UK), the ARC (Australia), the AAO, and the participating institutions. The GAMA website is http://www.gama-survey.org/ .

Funding for the SDSS and SDSS-II has been provided by the Alfred P. Sloan Foundation, the Participating Institutions,
the National Science Foundation, the U.S. Department of Energy, the National Aeronautics and Space
Administration, the Japanese Monbukagakusho, the Max
Planck Society, and the Higher Education Funding Council
for England. The SDSS Web Site is http://www.sdss.org/. The SDSS is managed by the Astrophysical Research
Consortium for the Participating Institutions. The Participating
Institutions are the American Museum of Natural
History, Astrophysical Institute Potsdam, University of
Basel, University of Cambridge, Case Western Reserve University,
University of Chicago, Drexel University, Fermilab,
the Institute for Advanced Study, the Japan Participation
Group, Johns Hopkins University, the Joint Institute for
Nuclear Astrophysics, the Kavli Institute for Particle Astrophysics
and Cosmology, the Korean Scientist Group, the
Chinese Academy of Sciences (LAMOST), Los Alamos National
Laboratory, the Max-Planck-Institute for Astronomy
(MPIA), the Max-Planck-Institute for Astrophysics (MPA),
New Mexico State University, Ohio State University, University
of Pittsburgh, University of Portsmouth, Princeton
University, the United States Naval Observatory, and the
University of Washington.

The VIKING survey is based on observations with ESO
Telescopes at the La Silla Paranal Observatory under the
programme ID 179.A-2004.

\bibliographystyle{mnras}
\bibliography{massfuncs}

\label{lastpage}
\end{document}